\documentclass[%
reprint,
superscriptaddress,
5preprint,
 amsmath,amssymb,
 aps,
]{revtex4-2}

\usepackage{graphicx}
\usepackage{dcolumn}
\usepackage{bm}

\usepackage{xcolor}
\usepackage[normalem]{ulem}

\usepackage[pagebackref=false]{hyperref}
\usepackage[capitalize]{cleveref}

\begin{document}

\author{A.~Rothstein}
\email{alexander.rothstein@rwth-aachen.de}
\affiliation{JARA-FIT and 2nd Institute of Physics, RWTH Aachen University, 52074 Aachen, Germany,~EU}%
\affiliation{Peter Gr\"unberg Institute  (PGI-9), Forschungszentrum J\"ulich GmbH, 52425 J\"ulich,~Germany,~EU}

\author{A.~Fischer} 
\affiliation{Institute for Theory of Statistical Physics, RWTH Aachen University, and JARA Fundamentals of Future Information Technology, 52062 Aachen, Germany}

\author{A.~Achtermann}
\affiliation{JARA-FIT and 2nd Institute of Physics, RWTH Aachen University, 52074 Aachen, Germany,~EU}%

\author{E.~Icking}
\affiliation{JARA-FIT and 2nd Institute of Physics, RWTH Aachen University, 52074 Aachen, Germany,~EU}%
\affiliation{Peter Gr\"unberg Institute  (PGI-9), Forschungszentrum J\"ulich GmbH, 52425 J\"ulich,~Germany,~EU}

\author{K.~Hecker}
\affiliation{JARA-FIT and 2nd Institute of Physics, RWTH Aachen University, 52074 Aachen, Germany,~EU}%
\affiliation{Peter Gr\"unberg Institute  (PGI-9), Forschungszentrum J\"ulich GmbH, 52425 J\"ulich,~Germany,~EU}

\author{L.~Banszerus}
\affiliation{JARA-FIT and 2nd Institute of Physics, RWTH Aachen University, 52074 Aachen, Germany,~EU}%
\affiliation{Peter Gr\"unberg Institute  (PGI-9), Forschungszentrum J\"ulich GmbH, 52425 J\"ulich,~Germany,~EU}

\author{M.~Otto}
\affiliation{AMO GmbH, Gesellschaft für Angewandte Mikro- und Optoelektronik, 52074 Aachen, Germany,~EU}

\author{S.~Trellenkamp}
\affiliation{Helmholtz Nano Facility, Forschungszentrum J\"ulich GmbH, 52425 J\"ulich,~Germany,~EU}

\author{F.~Lentz}
\affiliation{Helmholtz Nano Facility, Forschungszentrum J\"ulich GmbH, 52425 J\"ulich,~Germany,~EU}

\author{K.~Watanabe}
\affiliation{Research Center for Electronic and Optical Materials, 
National Institute for Materials Science, 1-1 Namiki, Tsukuba 305-0044, Japan}

\author{T.~Taniguchi}
\affiliation{Research Center for Materials Nanoarchitectonics, 
National Institute for Materials Science,  1-1 Namiki, Tsukuba 305-0044, Japan}

\author{B.~Beschoten}
\affiliation{JARA-FIT and 2nd Institute of Physics, RWTH Aachen University, 52074 Aachen, Germany,~EU}%

\author{R. J.~Dolleman}
\affiliation{JARA-FIT and 2nd Institute of Physics, RWTH Aachen University, 52074 Aachen, Germany,~EU}%

\author{D. M.~Kennes}
\affiliation{Institute for Theory of Statistical Physics, RWTH Aachen University, and JARA Fundamentals of Future Information Technology, 52062 Aachen, Germany,~EU}
\affiliation{Max Planck Institute for the Structure and Dynamics of Matter, Center for Free Electron Laser Science, 22761 Hamburg, Germany,~EU}

\author{C.~Stampfer}
\email{stampfer@physik.rwth-aachen.de}
\affiliation{JARA-FIT and 2nd Institute of Physics, RWTH Aachen University, 52074 Aachen, Germany,~EU}%
\affiliation{Peter Gr\"unberg Institute  (PGI-9), Forschungszentrum J\"ulich GmbH, 52425 J\"ulich,~Germany,~EU}%

\title{Gate-defined single-electron transistors in twisted bilayer graphene }

\date{\today}


\begin{abstract} 
\bf{Twisted bilayer graphene (tBLG) near the magic angle is a unique platform where the combination of topology and strong correlations gives rise to exotic electronic phases. 
These phases are gate-tunable and related to the presence of flat electronic bands, isolated by single-particle band gaps.
This enables gate-controlled charge confinement, essential for the operation of single-electron transistors (SETs), and allows to explore the interplay of confinement, electron interactions, band renormalisation and the moiré superlattice, potentially revealing key paradigms of strong correlations. 
Here, we present gate-defined SETs in near-magic-angle tBLG with well-tunable Coulomb blockade resonances. 
These SETs allow to study magnetic field-induced quantum oscillations in the density of states of the source-drain reservoirs, providing insight into gate-tunable Fermi surfaces of tBLG.
Comparison with tight-binding calculations highlights the importance of displacement-field-induced band renormalisation crucial for future advanced gate-tunable quantum devices and circuits in tBLG including e.g. quantum dots and Josephson junction arrays.}
\end{abstract}

\maketitle

\begin{figure*}[!t]
\centering
\includegraphics[draft=false,keepaspectratio=true,clip,width=0.99\linewidth]{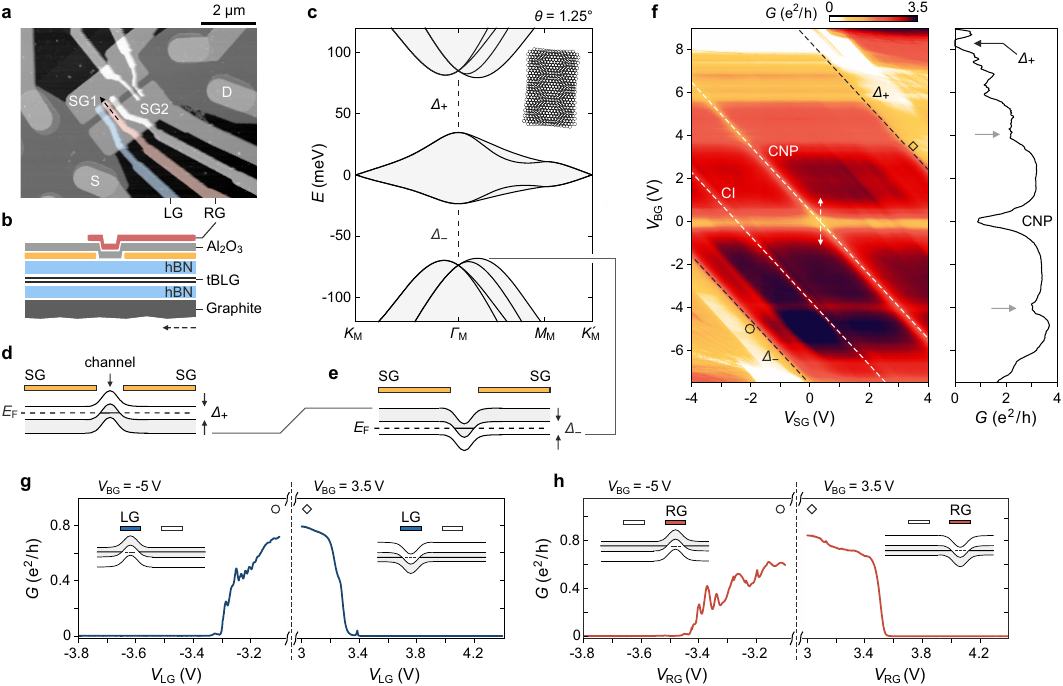}
\caption{\textbf{Device operation and tunable gate-defined conductive channels in tBLG.} 
(a) Scanning force microscopy image of the tBLG device. 
The relevant gates are highlighted. 
(b) Schematic cross section of the device along the dashed arrow in panel (a). 
(c) Tight-binding band structure calculation of tBLG at $\theta = 1.25^\circ$.
The single-particle band gaps $\Delta_{+}$ and $\Delta_{-}$emerging at full filling of the tBLG moir\'e unit cell are highlighted.
(d) Band edge profile along the device showing the formation of the split gate (SG) defined channel by tuning the Fermi level below the SGs into the gap $\Delta_+$ and in the channel into the flat band region. 
(e)~Same as in panel (d) but by using the gap $\Delta_-$ to define the channel.
(f) Two-terminal conductance $G = I/V_\mathrm{SD}$ as a function of $V_\mathrm{BG}$ and $V_\mathrm{SG} = V_\mathrm{SG1} = V_\mathrm{SG2}$ measured at $V_\mathrm{SD} = 100 \, \mathrm{\mu V}$. 
The line-cut (right panel) is taken at $V_\mathrm{SG} = 0.362 \, \mathrm{V}$ (white arrow in the conductance map) to compensate for intrinsic doping.
Gray arrows mark the half-filling correlated insulating states.
(g) Pinch-off traces for gate LG using the gap $\Delta_-$ [left panel: at $V_\mathrm{BG} = -5 \, \mathrm V$ and $V_\mathrm{SG1} = -2.14 \, \mathrm{V}$, $V_\mathrm{SG2} = -1.94 \, \mathrm{V}$ with $V_\mathrm{SD} = 100 \, \mathrm{\mu V}$; circle in panel (f)] and $\Delta_+$ [right panel: at $V_\mathrm{BG} = 3.5 \, \mathrm{V}$ and $V_\mathrm{SG1} = V_\mathrm{SG2} = 3.5\, \mathrm{V}$ with $V_\mathrm{SD} = 150 \, \mathrm{\mu V}$; rhombus in panel (f)].
Insets visualize the band edge bending along the SG-defined channel.
(h) Same as in panel (g) but for gate RG. 
\label{f1}}
\end{figure*}

Single-electron transistors (SETs)~\cite{Kastner1992Jul} are highly sensitive quantum electronic devices that exploit the control of individual charge carriers flowing through a small conducting island or a quantum dot. 
SETs have been realized in many different material systems (including van der Waals materials such as graphene~\cite{Ihn2010Mar} and 2D semiconductors~\cite{Zhang2017Oct}) and in particular with integrated quantum dots they permit the investigation of band structure properties, the study of electron-electron interactions~\cite{Tarucha1996Oct,Moller2021Dec}, as well as probing the density of states in nearby reservoirs~\cite{Banszerus2020Dec}.
All this makes the realisation of tunable SETs in twisted bilayer graphene interesting and promises additional avenues for studying its electronic properties.
Twisted bilayer graphene (tBLG) with a twist angle near $1^\circ$ not only has a rich phase diagram, including superconductivity~\cite{Cao2018Apr, Yankowitz2019Mar, Lu2019Oct, Wong2020Jun, Stepanov2020Jul, Saito2020Sep}, strange metal phases and correlated Chern insulators~\cite{Nuckolls2020Dec, Saito2021Apr, Wu2021Apr, Das2021Jun, Stepanov2021Nov}, but also offers sizeable single-particle band gaps at finite energy, making charge carrier confinement feasible.
The gate-tunability of these phases allowed the demonstration of built-in Josephson junctions~\cite{Cao2016Sep, Cao2018Apr, Lu2019Oct}, superconducting quantum interference devices~\cite{ Portoles2022Nov}, Aharonov-Bohm rings~\cite{Iwakiri2024Jan} and superconducting channels \cite{Zheng2024Mar}. 
However, well-controlled SETs and quantum dots - all of which require more complex gating structures - have not yet been demonstrated \cite{Tilak2021Jul}.

Here we demonstrate electrostatically controlled charge confinement and SET operation in tBLG based on an advanced gating structure.
By using a first set of top gates, we create a narrow conducting channel in the tBLG by exploiting the single-particle band gaps (on the order of several tens of meV) separating the flat bands from the remote conduction and valence band. 
An additional set of narrow gates across the channel allows local control of the band edge profile and enables the formation of isolated charge islands, very similar as in Bernal-stacked bilayer QD devices~\cite{Eich2018Jul,Banszerus2023Jun}.
This makes it possible to demonstrate sequential tunneling through flat-band charge islands and using the associated SET as a platform to perform magnetic field spectroscopy measurements, providing information about the Fermi surfaces in the nearby reservoirs and their sensitive dependence on out-of-plane electric fields, confirmed by tight-binding calculations.
In addition, our technology provides the basis for future studies of superconducting quantum point contacts, few electron quantum dots, Cooper pair splitters and related quantum devices in the tBLG.

\section*{Device operation}

The device consists of tBLG with a twist angle close to $\theta \approx 1.25^\circ$ (Supporting Section I), which is encapsulated in hexagonal boron nitride (hBN) and placed on a graphite flake acting as a back gate (BG). In addition, there are two metal top gate layers:
The first top gate layer consists of a pair of split gates (SGs), SG1 and SG2, which define an $\approx 3 \, \mathrm{\mu m} $ long and $\approx 200 \, \mathrm{nm}$ wide channel. 
The second top gate layer (separated from the first gate layer by 20~nm Al$_2$O$_3$) consists of a series of finger gates (FGs) $\approx 200$~nm wide and $\approx 120$~nm apart, placed above the channel so that the FGs cross the channel perpendicular to its orientation (see Methods for fabrication details). 
In \cref{f1}a,b we show a scanning force microscopy image of the device highlighting the left and right FG (LG and RG, respectively), and a schematic cross-section.

To best explain the device operation, we start by discussing the  band structure shown in \cref{f1}c highlighting the flat bands around zero energy and the single-particle band gaps $\Delta_{+}$ and $\Delta_{-}$ separating them from the remote bands.
In a first step, the BG and SG voltages, which locally tune the carrier density $n$, are adjusted so that the Fermi level $E_\text{F}$ in the region below the SGs lies in the band gap $\Delta_{+}$ (or $\Delta_{-}$), creating a narrow conducting channel defined by the SGs, as shown in \cref{f1}d (or \cref{f1}e).
In \cref{f1}f we show a $V_\mathrm{BG}$-$V_\mathrm{SG}$ conductance map, where the diagonal features (depending on  $V_\mathrm{BG}$ and $V_\mathrm{SG}$) originate from the areas below the SGs while the horizontal features (independent of $V_\mathrm{SG}$) originate from the areas not covered by the SGs.
We mark with dashed lines the diagonal features related to the charge neutrality point (CNP, $n=0$~cm$^{-2}$), the correlated insulating state (CI) at half filling of the fourfold valley and spin degenerate moir\'e bands, $n=-n_\mathrm{s}/2$ [with $n_\mathrm{s}=4/$(area of the moir\'e unit cell)] and the single-particle band gaps $\Delta_{\pm}$ at $\pm n_\mathrm{s}$ (see labels in \cref{f1}f).
The diagonal features exhibit a (negative) slope of $\approx1.5$ in good agreement with the ratio of the tBLG-BG and SG-tBLG capacitances given by $\approx 1.6$ (Supporting Section II and III). 

A conductive channel can now be defined by
e.g. setting the BG voltage to $V_\mathrm{BG} = -5$~V and the SG voltages (slightly asymmetric) to $V_\mathrm{SG1} = -2.14$~V and $V_\mathrm{SG2} = -1.94$~V (circle in \cref{f1}f). 
This tunes the Fermi level in the region below the SGs in the band gap $\Delta_{-}$, thereby defining a channel as illustrated in \cref{f1}e.
Next, we can use the individual FGs to completely suppress transport through the channel, as shown in the left panels of~\cref{f1}g,h.
By decreasing one of the FG voltages from $-3.2$ to $-3.6$~V we locally pull the band edges up in energy so that in the region below the LG (\cref{f1}g) or the RG (\cref{f1}h), the Fermi level is also in the band gap $\Delta_{-}$ (dashed lines in the left insets of \cref{f1}g,h).

As the device is gate-tunable, a similar result can be obtained by inverting the gate voltage polarities. This allows for the system to be tuned such that the channel is defined by placing $E_\text{F}$ in the gap $\Delta_+$ in the region below the SGs (see \cref{f1}d); for example, by setting $V_\mathrm{BG} = V_\mathrm{SG} = 3.5 \, \mathrm{V}$ (rhombus in \cref{f1}f).
In this case transport can be completely blocked by using the individual FGs to locally push the band edges down in energy, such that $E_\mathrm{F}$ resides in the gap $\Delta_+$ (shown by the right panels of~\cref{f1}g,h).

\begin{figure}[!t]
\centering
\includegraphics[draft=false,keepaspectratio=true,clip,width=1\linewidth]{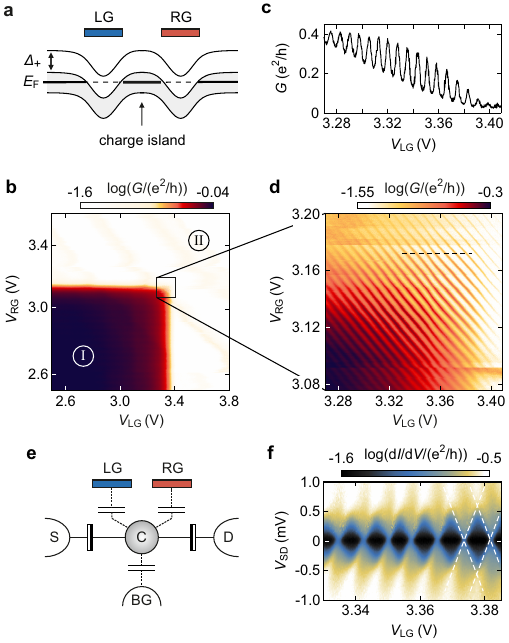}
\caption{\textbf{Formation of a gate-defined SET.}
(a) Schematic illustration of the band edge profile along the gate-defined channel showing the formation of charge confinement between LG and RG by using the gap $\Delta_+$.
(b) Large-range charge-stability diagram showing the conductance as a function of LG and RG measured at $V_\mathrm{BG} = 3.5$~V and $V_\mathrm{SG} = 3.5$~V (rhombus in \cref{f1}d) with $V_\mathrm{SD} = 100 \, \mathrm{\mu V}$.
(c)~Line-cut showing the onset of Coulomb oscillations along LG for fixed $V_\mathrm{RG} = 3.12$~V.
(d) Zoom-in into the area marked by the black rectangle in panel (b) showing a set of diagonal Coulomb resonances indicating single-electron transport through a charge island located in between the LG and RG.
(e) Schematic representation of the SET, with the charge island tunnel-coupled to the source (S) and drain (D) contacts and the corresponding gates highlighted.
(f) Finite bias spectroscopy measurements for fixed $V_\mathrm{RG} = 3.3301$~V [dashed line in panel (d)]. 
}   
\label{f2}
\end{figure}

Finally, SET operation can be achieved by using both FGs to modulate the band edge profile along the conductive channel as illustrated in \cref{f2}a.
\cref{f2}b shows a $V_\mathrm{LG}$-$V_\mathrm{RG}$ conductance map where we observe the transition from high conductance values (regime~I, where the FGs are tuned such that we have flat-band carrier transport) to strongly suppressed conductance (regime~II, where the FGs push the Fermi level into the gap $\Delta_+$).
Focusing on a small gate voltage range applied to the LG and RG (black square in \cref{f2}b and close-up in \cref{f2}d), we observe well-defined conductance oscillations (see also \cref{f2}c) that can be tuned equally well by the LG and RG (indicated by the slope of $\approx -1.08$ of the resonances in the $V_\mathrm{LG}$-$V_\mathrm{RG}$ plane).
These periodic conductance oscillations (with a constant spacing of $\Delta V_\mathrm{LG} \approx 7.9\,$~mV), which we identify as Coulomb resonances, are a hallmark of the presence of tunneling barriers below both FGs, which consequently electrostatically define a charge island between the left and right FG.
The occupation of this charge island, located in the flat bands, can be tuned by the capacitively coupled gates as illustrated in \cref{f2}e and confirmed by finite bias spectroscopy measurements revealing characteristic Coulomb diamonds (\cref{f2}f).

\section*{Highly tunable Coulomb resonances}
\begin{figure*}[t]
\centering
\includegraphics[draft=false,keepaspectratio=true,clip,width=1\linewidth]{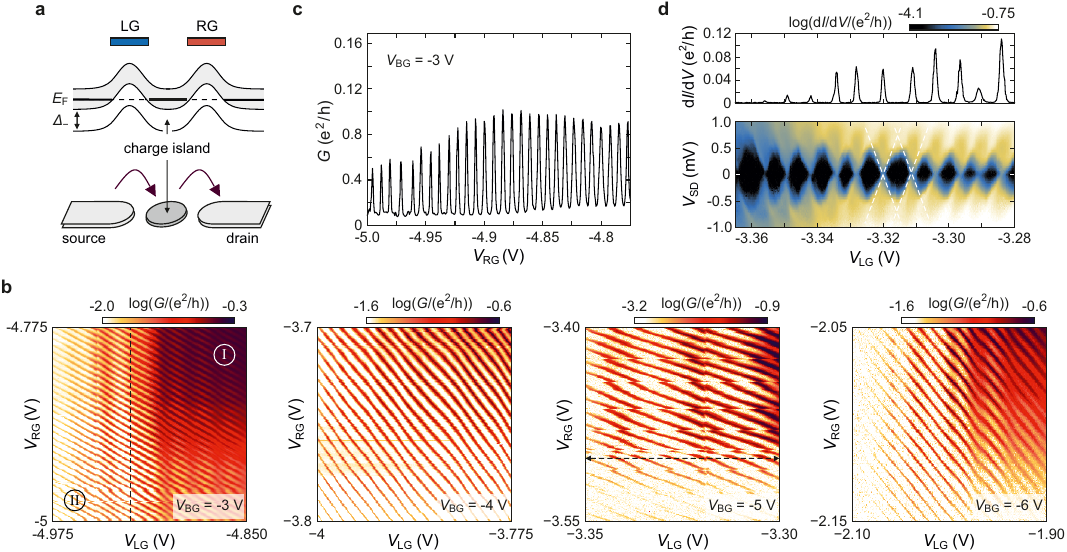}
\caption{
\textbf{Fully gate-tunable SET operation in the hole doped regime.}
(a) Schematic illustration showing the formation of a charge island between LG and RG by using the gap $\Delta_-$.
(b) Charge-stability diagrams showing diagonal Coulomb resonances for different combinations of BG and SG voltages highlighting the flexibility of the device. 
Measurements are taken at (i) $V_\mathrm{BG} = -3 \, \mathrm{V}$ and $V_\mathrm{SG} = -3.5 \, \mathrm{V}$ at $V_\mathrm{SD} = 150 \, \mu \mathrm{V}$.  
(ii) $V_\mathrm{BG} = -4 \, \mathrm{V}$ and $V_\mathrm{SG} = -3.23 \, \mathrm{V}$ at $V_\mathrm{SD} = 100 \, \mu \mathrm{V}$.
(iii) $V_\mathrm{SG} = -5 \, \mathrm{V}$ and (slightly asymmetric) split gate voltages of $V_\mathrm{SG1} = -2.14 \, \mathrm{V}$ and $V_\mathrm{SG2} = -1.94 \, \mathrm{V}$ at $V_\mathrm{SD} = 100 \, \mu \mathrm{V}$ (circle in Fig.~\ref{f1}f) and
(iv) $V_\mathrm{BG} = -6 \, \mathrm{V}$ and $V_\mathrm{SG} = -1.7 \, \mathrm{V}$ at $V_\mathrm{SD} = 100 \, \mu \mathrm{V}$. 
(c) Line-cut along the RG extracted from the left most panel in (b) for $V_\mathrm{BG} = -3\, \mathrm{V}$ at fixed $V_\mathrm{LG} = -4.925$~V showing periodic Coulomb oscillations with an approximate peak spacing of $\Delta V_\mathrm{RG} \approx 7.7$~mV.
(d)~Lower panel: Finite bias spectroscopy measurements as function of $V_\mathrm{LG}$ at fixed $V_\mathrm{BG} = -5\, \mathrm{V}$ and $V_\mathrm{RG} = -3.501$~V (see dashed line in central right panel in (b)). Upper panel: Zero bias Coulomb peak trace extracted from the Coulomb diamond measurement. Note the fact that complete current suppression can be achieved in this regime.
}   
\label{f3}
\end{figure*}

The high tunability of the device is demonstrated by inverting all gate voltages and using the gap $\Delta_-$ to form local tunneling barriers essential for the SET operation, as illustrated in \cref{f3}a.  
In \cref{f3}b, we show corresponding $V_\mathrm{LG}$-$V_\mathrm{RG}$ conductance maps for four different $V_{\rm BG}$ values, where $V_{\rm SG}$ is set to ensure that the Fermi level is well within the gap $\Delta_-$ in the tBLG region below the SGs, and the LG and RG are fine-tuned to form tunneling barriers below the FGs (see also Supporting Section IV).
In each case, we observe periodic Coulomb resonances (diagonal features), demonstrating the versatility of the device to create confinements using both single-particle band gaps of the tBLG electron system to realize a SET. 
Interestingly, we observe for all investigated regimes a nearly constant spacing of the Coulomb resonances of $\Delta V_\mathrm{LG}$ in the range of $(10.6 \pm 3.2)$~mV (see e.g. conductance trace shown in \cref{f2}c) and $\Delta V_\mathrm{RG}$ in the range of $(9.5 \pm 2)$~mV, both consistent with the value observed for the electron-doped regime. 
This, in conjunction with the well-developed diagonal slopes with values in the range of $\approx -0.6$ to $\approx -1.1$ (Supporting Section II) and the possibility to completely suppress the conductance in certain gate voltage regimes, which results in the formation of well-developed Coulomb peaks (\cref{f3}d), provides compelling evidence that the island is indeed formed consistently in the channel in between the two FGs.
From the characteristic spacing of the Coulomb oscillations $\Delta V_\mathrm{FG}$ we can estimate the capacitance between the island and the FGs given by $C_\mathrm{FG} = e/\Delta V_\mathrm{FG}$~\cite{vanderWiel2002Dec} resulting in $C_\mathrm{FG}$ in the range of $12$ to  $22$~$\mathrm{aF}$. 
Interestingly, this value is more than a magnitude smaller than the self-capacitance extracted from the charging energy of the islands, which is consistently found to be in the range of $\approx0.5$~meV (white dashed lines in \cref{f2}f and lower panel in \cref{f3}d), resulting in a self-capacitance of $C = e^2/E_\mathrm{c} \approx 320 \, \mathrm{aF}$.

Further insights can be gained by comparing this capacitance to the expected BG capacitance, which is expected to provide a dominant contribution. 
Assuming a simple plate capacitor with the area given by gate spacings ($A= 200 \times 120$~nm$^2$) we can estimate the BG capacitance by $C_\mathrm{BG} = \varepsilon_0 \varepsilon_\mathrm{r} A/d_\mathrm{b}$ resulting in $\approx 18$~aF (here we used $d_\mathrm{b} \approx 40$~nm and $\varepsilon_\mathrm{r} \approx 3.4$). This is also more than an order of magnitude smaller than the self-capacitance $C$ and interestingly comparable to the significantly more distant FG capacitances, strongly indicating that the area of the charge island extends substantially underneath the FGs  and (possibly the SGs) resulting in significantly larger island areas, consistent with the large self-capacitance. 
Moreover, it is expected that the capacitances to the source and drain reservoirs as well as to the SGs will play an important role in this system.

\begin{figure*}[!ht]
\centering
\includegraphics[draft=false,keepaspectratio=true,clip,width=1\linewidth]{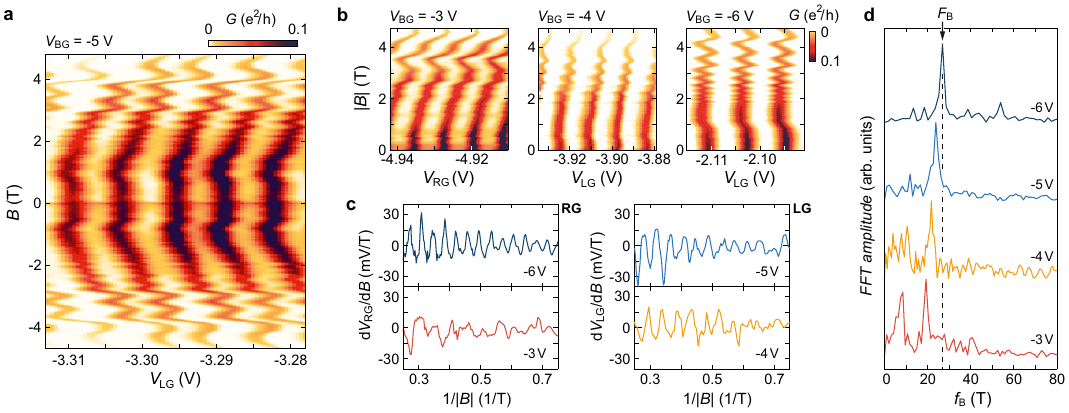}
\caption{\textbf{Out-of-plane magnetic field dependence of  Coulomb resonances at different BG voltages.}
(a)~Magnetospectroscopy of selected Coulomb resonances for $V_\mathrm{BG} = -5 \, \mathrm{V}$, $V_\mathrm{SG1} = -2.14 \, \mathrm{V}$ and $V_\mathrm{SG2} = -1.94 \, \mathrm{V}$ [see circle in \cref{f1}f and compare with \cref{f3}b] at $V_\mathrm{SD} = 220 \, \mu \mathrm{V}$ with $V_\mathrm{RG} = -3.472$~V. 
(b) Additional magnetospectroscopy data taken in the other BG regimes shown in \cref{f3}b. 
Measurements are taken at (i) $V_\mathrm{BG} = -3 \, \mathrm{V}$ and $V_\mathrm{SG} = -3.5 \, \mathrm{V}$ at $V_\mathrm{SD} = 220 \, \mu \mathrm{V}$ with $V_\mathrm{LG} = -4.925$~V. 
(ii) $V_\mathrm{BG} = -4 \, \mathrm{V}$ and $V_\mathrm{SG} = -3.23 \, \mathrm{V}$ at $V_\mathrm{SD} = 220 \, \mu \mathrm{V}$ with $V_\mathrm{RG} = -3.79$~V. 
(iii) $V_\mathrm{BG} = -6 \, \mathrm{V}$ and $V_\mathrm{SG} = -1.7 \, \mathrm{V}$ at $V_\mathrm{SD} = 220 \, \mu \mathrm{V}$ with $V_\mathrm{LG} = -2.017$~V. 
For the regimes (ii) and (iii) we performed a background subtraction to highlight the Coulomb resonances.
The conductance limits in regime (ii) are scaled by a factor $0.5$ for clarity.
(c) Oscillating behavior of $\mathrm{d} V_\mathrm{LG/RG}/\mathrm{d}B$ as a function of $1/|B|$ extracted from the tracked Coulomb resonances (i.e. peaks) in panels (a) and (b) averaged over the visible charge transitions. 
(g) Averaged frequency spectrum of the individual fast Fourier transforms (FFTs) of the oscillations shown in panel (c). For more information see text.
} 
\label{f4}
\end{figure*}

\section*{Detection of quantum oscillations}
\begin{figure}[!htbp]
\centering
\includegraphics[draft=false,keepaspectratio=true,clip,width=0.95\linewidth]{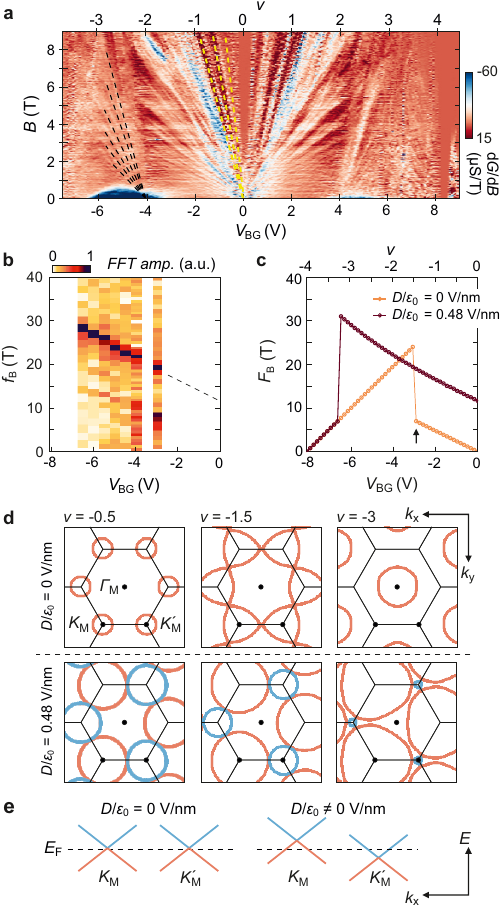}
\caption[]{\textbf{Gate-tunable Fermi surface.}
(a) $B$-field derivative of the bulk conductance as a function of $V_\mathrm{BG}$ (bottom axis) or filling factor $\nu$ (top axis) and out-of-plane magnetic field.
The dashed lines highlight the Landau levels emerging from CI at $\nu = -2$ (black) and the CNP at $\nu = 0$ (yellow).
(b) FFT of the $B$-field depnden quantum oscillations extracted for the different BG voltages (see Fig.~4d).
(c) Quantum oscillation frequency $F_B$ obtained from tight-binding calculations for displacement field strengths of $D/\varepsilon_0 = 0$~V/nm (orange data) and $D/\varepsilon_0 = 0.48$~V/nm (dark red data). 
Note, that for converting $\nu$ to $V_\mathrm{BG}$ we neglect here stray fields from other surrounding gates. 
(d) Contour plots of the Fermi surface for different filling factors (i.e. energies) around the $K_\mathrm{M}$ points without (top row) and with (bottom row) an applied displacement field.
(e) Schematic representation of the $D$-field induced energy shift. For more information see text.
}  
\label{f5}
\end{figure}
We now turn to the out-of-plane magnetic field ($B$-field) dependence of the Coulomb resonances. 
In \cref{f4}a,b we present Coulomb resonance measurements as a function of $V_\mathrm{LG}$ (or $V_\mathrm{RG}$) and $B$-field for different BG and SG voltage regimes as shown in \cref{f3}b.
The dependence of the Coulomb resonances on the $B$-field is generally quite complex. 
However, there are some common features, including (i) almost no $B$-field dependence for small $B$-fields (up to $0.3-1.2$~T depending on $V_\mathrm{BG}$), (ii) a general trend of shifting the Coulomb resonances towards lower energies, i.e. towards the CNP, consistent with previous work on graphene quantum dots~\cite{Guttinger2009Jul, Libisch2010Jun, Banszerus2020Oct} and (iii) the appearance of well-defined oscillations at higher $B$-fields, as also recently reported in Bernal-stacked bilayer graphene quantum dots~\cite{Banszerus2020Dec} and twist-angle domain-defined charge islands in tBLG~\cite{Dolleman2024Apr}.
Note that, as in Ref.~\cite{Dolleman2024Apr}, the spacing between the Coulomb resonances remains constant -- within our measurement resolution.

In the following we focus on the observed oscillations as they provide access to interesting information about the electronic structure of the gated flat-band system.
Note that these oscillations exhibit a periodicity inversely proportional to the applied $B$-field (see \cref{f4}c) and
are a consequence of the oscillating Fermi level in the source-drain reservoirs due to the formation of Landau levels~\cite{Banszerus2020Dec, Ihn2009Dec}. 
The frequency of the observed oscillations, which have the same origin as Shubnikov-de Haas oscillations, contains valuable information about the Fermi surface $\mathcal{S}$, which can be extracted using the Onsager relation $\mathcal S = 2 \pi e F_B/\hbar$, where $F_B$ is the frequency of the extracted $1/|B|$ oscillations.
To study the Fermi surface and its dependence on band filling, we conducted measurements of the $B$-field dependence of the Fermi surface for different values of $V_\mathrm{BG}$, as demonstrated in  \cref{f4}a,b.
In \cref{f4}c, we present the corresponding $B$-field derivatives of the gate voltage peak positions, plotted as a function of  inverse $B$-field.
Each trace represents an average taken over at least three Coulomb resonances (Supporting Section V).
The corresponding fast Fourier transforms (FFTs) are shown in \cref{f4}d, which highlight that there is indeed one dominant $1/|B|$ frequency, $F_B$. This frequency shifts to lower values when  $V_\mathrm{BG}$ is decreased.

Interestingly, there are some at first sight unexpected differences when comparing this data with the bulk magneto-conductance measurements for which the SGs and FGs are tuned such that the entire tBLG area is uniformly conductive (Supporting Section I).
In \cref{f5}a we show such a bulk measurement, specifically $\mathrm d G/\mathrm dB$ as function of $V_\mathrm{BG}$ and $B$-field and find the well expected emerging Landau fan from $V_\mathrm{BG} = 0$~V (see yellow dashed lines) which is connected to a Fermi surface area $\mathcal S$ that vanishes at the CNP (filling factor $\nu = 0$) in good agreement with earlier work~\cite{Cao2016Sep}.
Moreover, we observe additional fans emerging from the CI states at $\nu = \pm 2$ (see e.g. black dashed lines highlighting the fan at $\nu = -2$) implying that $\mathcal S$ should exhibit a minimum (if not vanishing) at these points.
Let us now turn to the BG voltage dependent Fermi surface, i.e. $F_B$, extracted from the Coulomb resonance measurements discussed in \cref{f4} and summarised in \cref{f5}b.
Here we observe the expected linear tuning of the main $1/|B|$ frequency component, but there is no minimum (or zero) at $V_\mathrm{BG} \approx -4$~V, where we would expect the CI state at $\nu = -2$ (compare to \cref{f5}a).
Furthermore, by extending a linear slope to $V_\mathrm{BG} = 0 \, \mathrm{V}$ (dashed line in \cref{f5}b), we conclude that there is most likely a finite Fermi surface at $V_\mathrm{BG}=0$~V, in contrast to the measurements shown in \cref{f5}a, and in contrast to large-scale tight-binding calculations (see orange trace in \cref{f5}c)~\cite{Bistritzer2011Jul, LopesdosSantos2007Dec, Shallcross2010Apr, SuarezMorell2010Sep, Koshino2018Sep}.
Note that our tight-binding calculations (details in Supporting Section VII) are in good agreement with previous works~\cite{Cao2016Sep,Kim2016Aug} and our bulk measurements (\cref{f5}a).
However, when compared with the data obtained from Coulomb resonance magnetospectroscopy (\cref{f5}b), the tight-binding calculation reveals an additional difference: the expected Lifshitz transition at $\nu \approx -1.5$ (black arrow in \cref{f5}c -- expected at $V_\mathrm{BG}\approx -3$~V), which leads to a sign reversal of the slope of $F_\mathrm{B}$ vs. $V_\mathrm{BG}$ is completely absent in the data shown in \cref{f5}b.
This Lifshitz transition, which is related to a van Hove singularity in the density of states, describes the transition from Fermi surfaces associated with valley-exclusive hole-like orbits around the moir\'e Dirac points $K_\mathrm{M}^{(\prime)}$ to Fermi surfaces associated with electron-like orbits centred around $\Gamma_\mathrm{M}$, as shown in \cref{f5}d (upper row).

The absence of the Lifshitz transition and the indication of a finite Fermi surface at the expected CNP can in principal be explained by the presence of (i) an out-of-plane electric field \cite{Dale2023Aug} or (ii) heterostrain \cite{Bi2019Jul, Wang2023Aug}, both having sizeable impact on the band structure of tBLG near the magic angle.
From the bulk measurements (\cref{f5}a), in particular from the fact that there is only one Landau fan emerging from the CNP, we rule out heterostrain as the dominant cause of the finite electron/hole Fermi surfaces at the CNP \cite{Piccinini2022Jul}, making out-of-plane displacement fields ($D$-fields) the prime candidate responsible for the band deformations.
Such $D$-fields are likely present due to the multiple gate layers with different potentials, i.e. gate voltages. Crucially, a finite $D$-field leads to an energy difference between the $p_\mathrm{z}$-orbitals in the upper and lower graphene layer of the tBLG, which creates an energetic asymmetry between the cones (band touching points) at the moir\'e Dirac $K_\mathrm{M}$ and $K_\mathrm{M}^{\prime}$ points.
This energetic offset induced by the displacement field (\cref{f5}e) results (i) in a finite Fermi surface at the CNP, where both conductance and valence band carriers are  contributing, and (ii) in a shift of the Lifshitz transition to higher hole filling.

In \cref{f5}c (dark red data points) and \cref{f5}d (lower row), we present the corresponding tight-binding calculations for an applied displacement field of $D/\varepsilon_0 = 0.48$ V/nm, which is a reasonable value for the applied gate voltages (Supporting Section IX).
The energetic offset of the Dirac cones at the $K_\mathrm{M}$ and $K_\mathrm{M}^{\prime}$ points results in a finite Fermi surface (i.e. in a finite $F_B$) at the CNP in quantitative agreement with the experiment (compare \cref{f5}b,c).
Near the CNP, the Fermi surface within a single valley is made up of valence band (orange) and conductance band (blue) pockets as shown in the lower left panel in \cref{f5}d.
With increasing hole filling the imbalance between the valance band and conduction band pockets increases (\cref{f5}d) with the overall consequence that the Lifshitz transition is shifted to values below  $\nu < -3$ in agreement with the experimental data (compare \cref{f5}b,c).
Moreover, we expect that the presence of strong displacement fields in the gate-defined device geometry significantly screens the long-ranged Coulomb interaction~\cite{Goodwin2020screening,Stepanov2020Jul,Liu2021Mar} suppressing the formation of the correlated insulating states at $\nu= -2$, in agreement with our experimental data (\cref{f5}b) and in line with theoretical predictions of interaction-induced band renormalisation \cite{Guinea2018,goodwin2020hartree,Rademaker2019,Cea2019} (Supporting Section VIII). 
Thus, our findings stress the significant impact of out-of-plane electrical fields on the band structure of near magic-angle tBLG.

\section*{Discussion and conclusions}

In conclusion, our measurements show that we can implement gate-defined SET in tBLG and use them as a platform to gain insights in band deformations of (near) magic angle tBLG in the presence of out-of-plane electric fields.
Essentially, our extracted data fill the gap between the large-angle regime ($\theta \gg 1^\circ$), which is characterized by large bandwidths and comparatively little sensitivity to external electrical fields~\cite{Piccinini2022Jul,Cao2016Sep} and the small-angle regime ($\theta \ll 1^\circ$), where the presence of electric fields causes the formation of topological helical networks~\cite{rickhaus2018transport,san2013helical,huang2018topologically}.
An important observation is the absence of the correlated insulating state in \cref{f5}b, while it is prominently present in the bulk transport data shown in \cref{f5}a. 

Given the recent theoretical and experimental study on the importance of long-ranged Coulomb interactions in the formation of correlated states in (near) magic-angle tBLG~\cite{nuckolls2023quantum,Stepanov2020Jul,Liu2021Mar}, we emphasize the flexibility of our device architecture to study electric field-induced screening of electronic interactions.
In this respect, future experiments even closer to the magic angle could provide much-needed answers to a number of pressing questions related to correlated phenomena such as superconductivity and strange metal phases and their interplay with topology.
In addition, the technology presented can enable the development of complex quantum devices and circuits based on tBLG, ranging from coupled quantum dots, Cooper-pair splitters, chains of Josephson junction-based SQUIDs to single-photon detector arrays. \\

\textbf{Acknowledgements} 
This project has received funding from the European Research Council (ERC) under grant agreement No. 820254, the Deutsche Forschungsgemeinschaft (DFG, German Research Foundation) through SPP 2244 (Project No. 535377524) and under Germany’s Excellence Strategy - Cluster of Excellence Matter and Light for Quantum Computing (ML4Q) EXC 2004/1 - 390534769, the FLAG-ERA grants No. 437214324 TATTOOS, No. 471733165 PhotoTBG, No. 534269806 ThinQ by the Deutsche Forschungsgemeinschaft
(DFG, German Research Foundation), and by the Helmholtz Nano Facility~\cite{Albrecht2017May}. K.W. and T.T. acknowledge support from the JSPS KAKENHI (Grant Numbers 21H05233 and 23H02052) and World Premier International Research Center Initiative (WPI), MEXT, Japan. \\

\textbf{Author contributions} 
C.S. and A.R. conceived the experiment.
A.R. built the device and performed the measurements. 
A.F. performed the theoretical simulations. 
A.R. and A.A. performed the data analysis under the supervision of C.S.. 
E.I., K.H and L.B. provided assistance with the measurement setup. 
M.O. performed the ALD step. 
S.T. and F.L. performed the electron beam lithography. 
K.W. and T.T. supplied the hBN crystals. 
A.R., A.F., R.J.D., D.M.K., B.B. and C.S. discussed the experimental and theoretical data and wrote the manuscript. \\

\textbf{Data availability} 
The data supporting the findings of this study are available in a Zenodo repository under [insert DOI link here].

\section*{Methods}
\subsection*{Device Fabrication}
The device consists of a hBN-tBLG-hBN heterostructure (thicknesses of the top and bottom hBN are measured via atomic force microscopy and approximately given by $d_\mathrm{t} \approx 25$~nm and $d_\mathrm{b} \approx 40$~nm, respectively) with a graphite BG fabricated by a 'laser-cut-and-stack' technique following a standard PDMS/PC transfer process \cite{Kim2016Mar, Park2021Apr, Rothstein2024Apr}. 
After one-dimensional side contacting (Cr/Au, $5/50$~nm) of the tBLG \cite{Wang2013Nov}, we fabricate the first gating layer containing the SGs (Cr/Au, $5/50$~nm) on top of the upper hBN layer. 
These gates are $3$~$\mathrm{\mu}$m wide and are separated by $200$~nm, defining a quasi one-dimensional channel [see scanning force microscopy image in Fig.~1a]. 
Following the SG metallisation, we evaporate a hard mask (Al, $60$~nm) to subsequently structure the mesa by means of reactive ion etching ($\mathrm{SF_6/O_2}$). 
In the next step, we use atomic layer deposition (ALD) of alumina ($\mathrm{Al_2O_3}$) to make a $15$~nm thick insulating layer separating the SGs from the FG layer.
Finally, we structure the FGs (Cr/Au, $5/170$~nm), which intersect the previously defined channel. 
The three FGs are approximately $200$~nm wide and are separated by around $120$~nm. 
A schematic cross section along the device is shown in \cref{f1}b

\subsection*{Measurement Setup}
All measurements presented in this manuscript were performed in a ${}^3\mathrm{He}/{}^4\mathrm{He}$ (Oxford Instruments, Triton 200) dilution refrigerator at a temperature below $100$~mK.
In a two-terminal setup we apply a fixed bias voltage over a home-built IV-converter via the source-drain contacts highlighted in \cref{f1}a and parallel measure the current with a digital multimeter (Agilent 34401). 
For the bias spectroscopy measurements shown in \cref{f2}f and \cref{f3}d we used a lock-in amplifier (Stanford Research SR830) at a frequency of $14.5$~Hz in a two-terminal setup.

\end{document}


\author{A.~Rothstein}
\email{alexander.rothstein@rwth-aachen.de}
\affiliation{JARA-FIT and 2nd Institute of Physics, RWTH Aachen University, 52074 Aachen, Germany,~EU}%
\affiliation{Peter Gr\"unberg Institute  (PGI-9), Forschungszentrum J\"ulich GmbH, 52425 J\"ulich,~Germany,~EU}

\author{A.~Fischer} 
\affiliation{Institute for Theory of Statistical Physics, RWTH Aachen University, and JARA Fundamentals of Future Information Technology, 52062 Aachen, Germany}

\author{A.~Achtermann}
\affiliation{JARA-FIT and 2nd Institute of Physics, RWTH Aachen University, 52074 Aachen, Germany,~EU}%

\author{E.~Icking}
\affiliation{JARA-FIT and 2nd Institute of Physics, RWTH Aachen University, 52074 Aachen, Germany,~EU}%
\affiliation{Peter Gr\"unberg Institute  (PGI-9), Forschungszentrum J\"ulich GmbH, 52425 J\"ulich,~Germany,~EU}

\author{K.~Hecker}
\affiliation{JARA-FIT and 2nd Institute of Physics, RWTH Aachen University, 52074 Aachen, Germany,~EU}%
\affiliation{Peter Gr\"unberg Institute  (PGI-9), Forschungszentrum J\"ulich GmbH, 52425 J\"ulich,~Germany,~EU}

\author{L.~Banszerus}
\affiliation{JARA-FIT and 2nd Institute of Physics, RWTH Aachen University, 52074 Aachen, Germany,~EU}%
\affiliation{Peter Gr\"unberg Institute  (PGI-9), Forschungszentrum J\"ulich GmbH, 52425 J\"ulich,~Germany,~EU}

\author{M.~Otto}
\affiliation{AMO GmbH, Gesellschaft für Angewandte Mikro- und Optoelektronik, 52074 Aachen, Germany,~EU}

\author{S.~Trellenkamp}
\affiliation{Helmholtz Nano Facility, Forschungszentrum J\"ulich GmbH, 52425 J\"ulich,~Germany,~EU}

\author{F.~Lentz}
\affiliation{Helmholtz Nano Facility, Forschungszentrum J\"ulich GmbH, 52425 J\"ulich,~Germany,~EU}

\author{K.~Watanabe}
\affiliation{Research Center for Electronic and Optical Materials, 
National Institute for Materials Science, 1-1 Namiki, Tsukuba 305-0044, Japan}

\author{T.~Taniguchi}
\affiliation{Research Center for Materials Nanoarchitectonics, 
National Institute for Materials Science,  1-1 Namiki, Tsukuba 305-0044, Japan}

\author{B.~Beschoten}
\affiliation{JARA-FIT and 2nd Institute of Physics, RWTH Aachen University, 52074 Aachen, Germany,~EU}%

\author{R. J.~Dolleman}
\affiliation{JARA-FIT and 2nd Institute of Physics, RWTH Aachen University, 52074 Aachen, Germany,~EU}%

\author{D. M.~Kennes}
\affiliation{Institute for Theory of Statistical Physics, RWTH Aachen University, and JARA Fundamentals of Future Information Technology, 52062 Aachen, Germany,~EU}
\affiliation{Max Planck Institute for the Structure and Dynamics of Matter, Center for Free Electron Laser Science, 22761 Hamburg, Germany,~EU}

\author{C.~Stampfer}
\email{stampfer@physik.rwth-aachen.de}
\affiliation{JARA-FIT and 2nd Institute of Physics, RWTH Aachen University, 52074 Aachen, Germany,~EU}%
\affiliation{Peter Gr\"unberg Institute  (PGI-9), Forschungszentrum J\"ulich GmbH, 52425 J\"ulich,~Germany,~EU}%

\title{Supporting Information - Gate-defined single-electron transistors in twisted bilayer graphene}

\date{\today}


\maketitle

\renewcommand{\thefigure}{S\arabic{figure}}
\setcounter{figure}{0}
\newpage

\section{Twist angle estimation}
\begin{figure}[!htbp]
\centering
\includegraphics[draft=false,keepaspectratio=true,clip,width=0.99\linewidth]{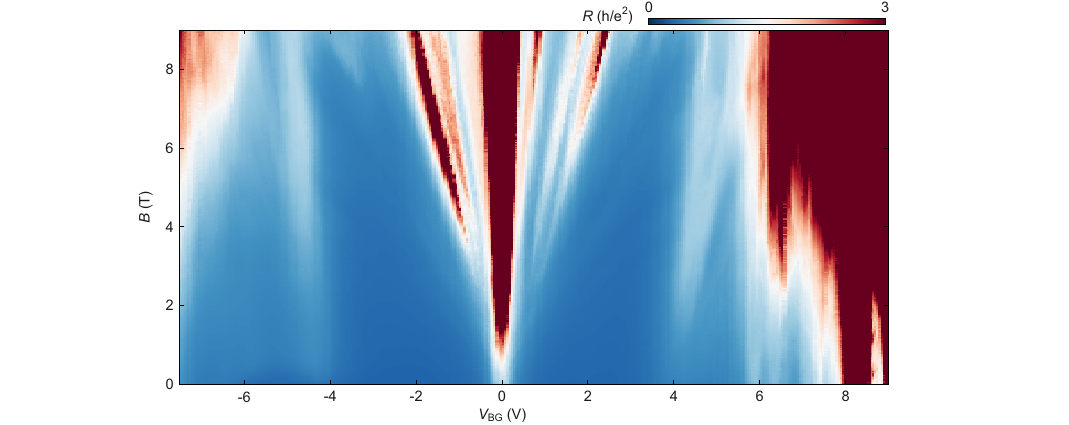}
\caption{\textbf{Bulk magnetoresistance of the device.}
Landau fan diagram showing the resistance $R$ as a function of $V_\mathrm{BG}$ and out-of-plane magnetic field $B$ recorded at $V_\mathrm{SD} = 200 \, \mu  V$ and $V_\mathrm{SG}=0.362$~V to compensate for intrinsic doping (see white double arrow in Fig.~1d of the main manuscript).
} \label{Sf0}
\end{figure}

We estimate the twist angle of the stacked tBLG by analysing the bulk magnetotransport measurements as shown in \cref{Sf0}, where we tune the charge carrier density in the tBLG with the graphite BG. 
We first extract the gate lever arm of the BG by least-square fitting of resistance minima of the visible Landau levels emerging from charge neutrality. The lever arm is then accesible via
\begin{align}
    B_\mathrm{LL} = \frac{h}{\nu_\mathrm{LL} e}\alpha_\mathrm{BG}V_\mathrm{BG} + \mathrm{const.},
\end{align}
where $\nu_\mathrm{LL} = \pm 2, \pm 4$ denotes the observed Landau level degeneracy.
This proceeding results in a value of $\alpha_\mathrm{BG} = (4.53 \pm 0.46) \times 10^{11} \, \mathrm{V^{-1}cm^{-2}}$, which is in reasonable agreement with the theoretical expected value of $\alpha_\mathrm{BG} = 4.70 \times 10^{11} \, \mathrm{V^{-1}cm^{-2}}$ calculated via a simple plate capacitor model $\alpha_\mathrm{BG} = \varepsilon_0 \varepsilon_\mathrm{hBN}/(ed_\mathrm b)$. 
Here, we used $\varepsilon_\mathrm{hBN} = 3.4$ \cite{Pierret2022Jun, Laturia2018Mar} and a bottom hBN thickness of $d_\mathrm{b} \approx 40$~nm.
To extract the superlattice carrier density $n_\mathrm{s}$ we convert the applied BG voltage via $n = \alpha_\mathrm{BG}(V_\mathrm{BG} - V_\mathrm{BG,off})$ to a charge carrier density.
We estimate the offset to be $V_\mathrm{BG,off} = 0\, \mathrm{V}$ since the Landau fan emerging from charge neutrality is located symmetrically around $V_\mathrm{BG} = 0 \, \mathrm{V}$.
Using the emerging fan at $V_\mathrm{BG} \approx 4.05$~V (which we identify with the correlated gap at half-filling) we estimate the superlattice density to be $n_\mathrm{s} = (3.67 \pm 0.38) \times 10^{12}\, \mathrm{cm^{-2}}$. 
The twist angle is then directly calculated via $\theta = [\sqrt{3}n_\mathrm{s}^2/8]^{1/2}a$ (with $a = 0.246\, \mathrm{nm}$ being the graphene lattice constant) which gives a value of $\theta = 1.26^\circ \pm 0.07^\circ$.

\newpage 

\section{Estimates of capacitances}
\paragraph{Plate capacitor model for the tBLG-BG and SG-tBLG capacitances}
We estimate the ratio of the tBLG-BG and SG-tBLG coupling by using a simple plate capacitor model $C = \varepsilon_0 \varepsilon_r A/d$ with area $A$ and gate distance $d$ (given by the hBN thicknesses).
We thus receive for the ratio $C_\mathrm{BG}/C_\mathrm{SG} = d_\mathrm t/d_\mathrm b \approx 1.5$ which is in good agreement with the experimentally extracted value of $\approx 1.6$.
\paragraph{Finger gate capacitances extracted from charge-stability diagrams}
For the charge confinements we extract the capacitive coupling of the finger gates to the charge islands directly from the spacing of the charging lines in the charge stability diagrams via $C = e/\Delta V_\mathrm{FG}$ as shown in \cref{Tab_cap}.

\begin{table}[h]
\centering
\def\arraystretch{1.2}
\begin{tabular}{c|c|c|c|c|cc}
$V_\mathrm{BG} \, \mathrm{(V)}$ & $\Delta V_\mathrm{LG} \, \mathrm{(mV)}$ & $C_\mathrm{LG} = e/\Delta V_\mathrm{LG} \, \mathrm{(aF)}$ & $\Delta V_\mathrm{RG} \, \mathrm{(mV)}$ & $C_\mathrm{RG}= e/ \Delta V_\mathrm{RG} \, \mathrm{(aF)}$ & slope   &  \\ \cline{1-6}
$-2.89$                         & $8.4$                                   & $19.1$                                                    & $7.5$                                   & $21.4$                                                    & $-0.89$ &  \\
$-3$                            & $7.4$                                   & $21.7$                                                    & $7.7$                                   & $20.8$                                                    & $-1.04$ &  \\
$-4$                            & $13.8$                                  & $11.6$                                                    & $7.6$                                   & $21.1$                                                    & $-0.55$ &  \\
$-4.5$                          & $13.0$                                  & $12.3$                                                    & $8.0$                                   & $20.0$                                                    & $-0.62$ &  \\
$-5$                            & $8.3$                                   & $19.3$                                                    & $8.9$                                   & $18.0$                                                    & $-1.07$ &  \\
$-5.5$                          & $13.8$                                  & $11.6$                                                    & $7.9$                                   & $20.3$                                                    & $-0.57$ &  \\
$-6$                            & $12.6$                                  & $12.7$                                                    & $7.9$                                   & $20.3$                                                    & $-0.63$ &  \\
$-6.5$                          & $12.2$                                  & $13.1$                                                    & $8.3$                                   & $19.3$                                                    & $-0.68$ &  \\
$-7$                            & $12.7$                                  & $12.6$                                                    & $11.5$                                  & $13.9$                                                    & $-0.91$ &  \\
$+3.5$                          & $7.9$                                   & $20.3$                                                    & $8.5$                                   & $18.8$                                                    & $-1.08$ & 
\end{tabular}
\caption{Finger gate voltage spacings between the Coulomb charging lines and associated gate capacitance extracted from the individual charge-stability diagrams.}
\label{Tab_cap}
\end{table}

\newpage 

\section{Individual BG vs. SG conductance maps}

\begin{figure}[!htbp]
\centering
\includegraphics[draft=false,keepaspectratio=true,clip,width=0.99\linewidth]{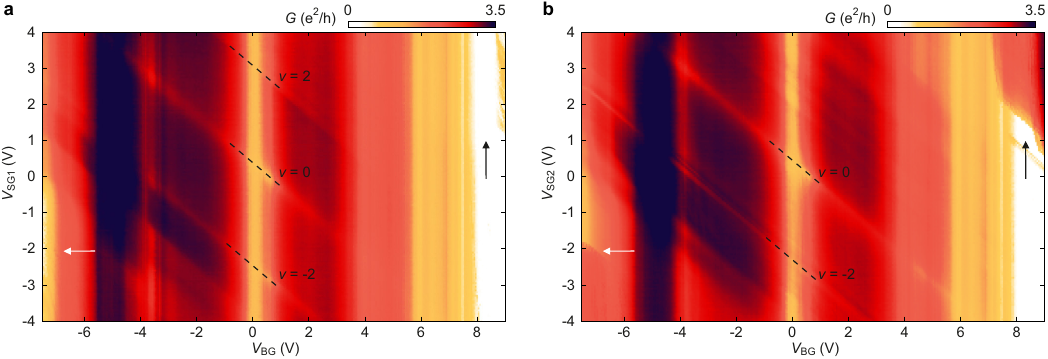}
\caption{\textbf{Individual split gate vs. back gate maps.}
(a) Two-terminal conductance as a function of $V_\mathrm{SG1}$ and $V_\mathrm{BG}$ recorded at $V_\mathrm{SD} = 100 \, \mathrm{\mu V}$. 
The voltages of all other gates were kept at zero. (b) Same as in (a) but as a function of $V_\mathrm{SG2}$ and $V_\mathrm{BG}$.} \label{Sf1}
\end{figure}

To investigate the twist angle homogeneity below the individual SGs and to characterize their functionality, we measure $G$ along the device as a function of the BG and SG1 (or SG2) while SG2 (or SG1) is kept at $0 \, \mathrm{V}$. 
The results are plotted in Fig.~\ref{Sf1}a,b. 
We observe slight deviations between the two maps which we attribute to small twist angle variation under areas covered by the SGs \cite{Uri2020May, Schapers2022Jul}. 
For example, we observe a diagonal line which is corresponding to the $\nu = 2$ correlated insulator in Fig.~\ref{Sf1}a which is almost completely absent in Fig.~\ref{Sf1}b (see dashed lines in both panels as a guide to the eye). 
Furthermore, we observe a breaking of the electron-doped band insulator in Fig.~\ref{Sf1}b for $V_\mathrm{BG} > 8\, \mathrm{V}$ and $V_\mathrm{SG2}> 1 \, \mathrm{V}$ which does not appear for $V_\mathrm{SG1}> 1 \, \mathrm{V}$ (see black arrows). 
On the hole-doped side, the band insulator is more pronounced below SG1 (see white arrows).

\newpage
\section{Additional Charge-stability diagrams}
\begin{figure}[!htbp]
\centering
\includegraphics[draft=false,keepaspectratio=true,clip,width=0.99\linewidth]{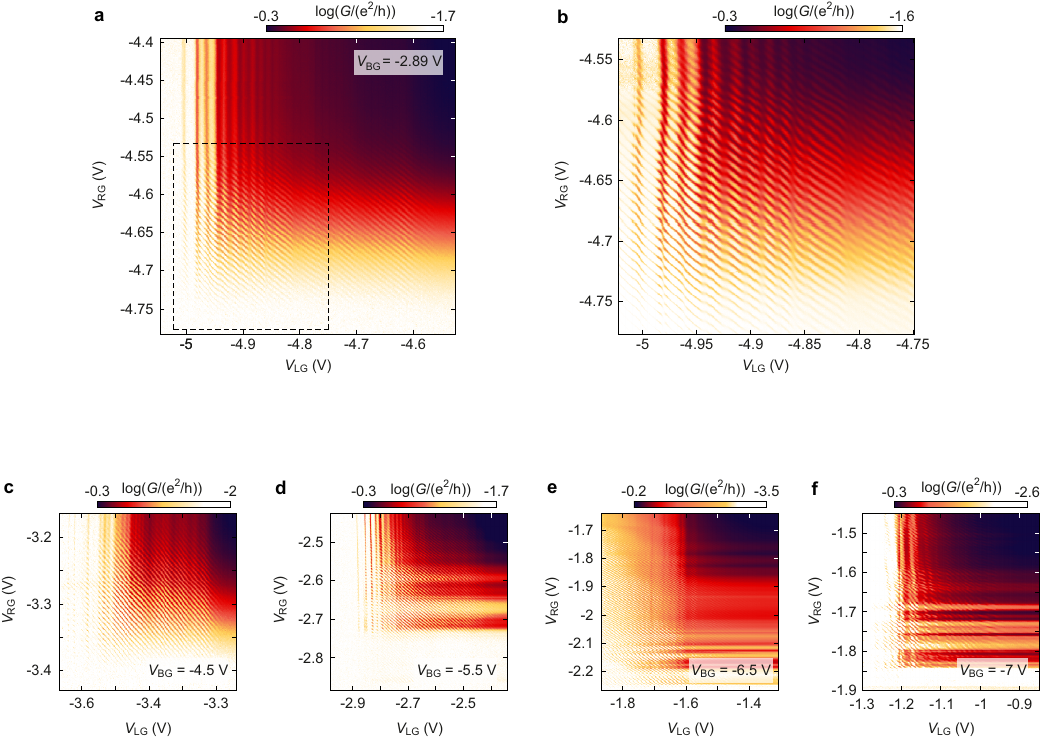}
\caption{\textbf{Additional charge-stability diagrams recorded in the hole-doped regime.}
Logarithmic two-terminal conductance as a function of the LG and the RG. 
(a) $V_\mathrm{BG} = -2.89$~V, $V_\mathrm{SG} = -3.85$~V at $V_\mathrm{SD} = 150\, \mathrm{\mu V}$. 
(b) Zoom-in into the area marked by the dashed rectangle in panel (a). 
(c) $V_\mathrm{BG} = -4.5$~V, $V_\mathrm{SG} = -2.82$~V at $V_\mathrm{SD} = 100\, \mathrm{\mu V}$. 
(d) $V_\mathrm{BG} = -5.5$~V, $V_\mathrm{SG} = -1.86$~V at $V_\mathrm{SD} = 150 \, \mathrm{\mu V}$. 
(e) $V_\mathrm{BG} = -6.5$~V, $V_\mathrm{SG} = -1.1$~V at $V_\mathrm{SD} = 150 \, \mathrm{\mu V}$. 
(f) $V_\mathrm{BG} = -7$~V, $V_\mathrm{SG} =-0.75$~V at $V_\mathrm{SD} = 100 \, \mathrm{\mu V}$.} \label{Sf2}
\end{figure}

We realised the charge confinements for further additional BG and split SG combinations. 
In \cref{Sf2} we show charge-stability diagrams for five combinations which are not given in the main manuscript. 

\newpage
\section{Additional magnetospectroscopy regimes}
\begin{figure}[!htbp]
\centering
\includegraphics[draft=false,keepaspectratio=true,clip,width=0.99\linewidth]{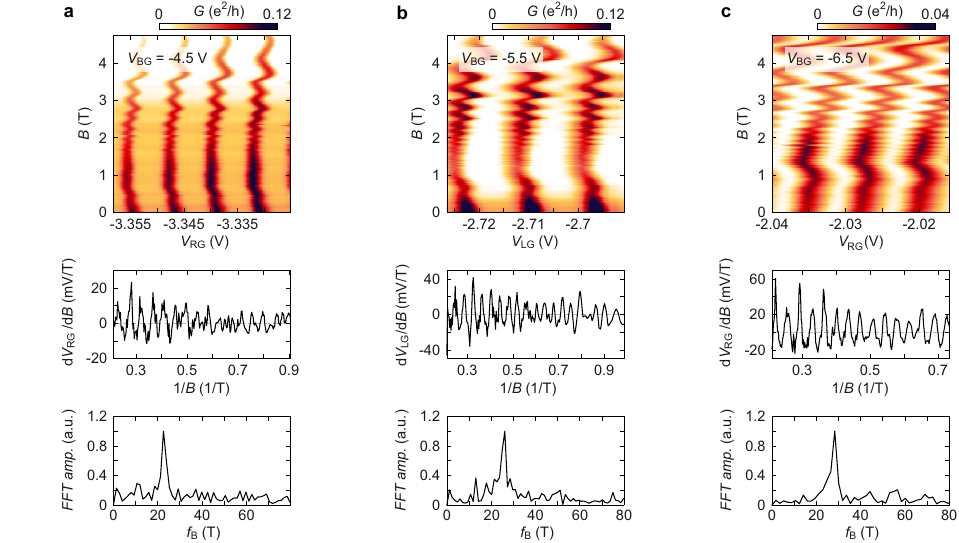}
\caption{\textbf{Additional magnetic field dependence of Coulomb resonances.} 
Magnetospectroscopy of selected Coulomb resonances along the RG (or LG) for the regimes shown in \cref{Sf2}c-e (top row). 
The shift of the Coulomb resonances (i.e. peaks) is tracked and averaged to extract the
oscillating behaviour of $\mathrm{d}V_\mathrm{LG/RG}/\mathrm{d}B$ as a function of 1/B (middle row). 
The frequency spectrum of the FFT shows a prominent peak which shifts with increasing BG voltage (bottom row). 
(a) $V_\mathrm{BG} = -4.5$~V, $V_\mathrm{SG} = -2.82$~V and $V_\mathrm{LG} = -3.5321$~V at $V_\mathrm{SD} = 220 \, \mathrm{\mu V}$.
(b) $V_\mathrm{BG} = -5.5$~V, $V_\mathrm{SG} = -1.86$~V and $V_\mathrm{RG} = -2.6857$~V at $V_\mathrm{SD} = 220 \, \mathrm{\mu V}$.
(c) $V_\mathrm{BG} = -6.5$~V, $V_\mathrm{SG} = -1.1$~V and $V_\mathrm{LG} = -1.735$~V at $V_\mathrm{SD} = 250 \, \mathrm{\mu V}$.} \label{Sf3}
\end{figure}

In \cref{Sf3} we show additional magnetospectroscopy measurements taken at BG voltages of $V_\mathrm{BG} = -4.5$~V, $V_\mathrm{BG} = -5.5$~V and $V_\mathrm{BG} = -6.5$~V. 
We observe an analogous characteristic of the out-of-plane magnetic field tuning as for the previously discussed regimes. 

\newpage
\section{Frequency analysis of the magnetospectroscopy data}
\begin{figure}[!htbp]
\centering
\includegraphics[draft=false,keepaspectratio=true,clip,width=0.99\linewidth]{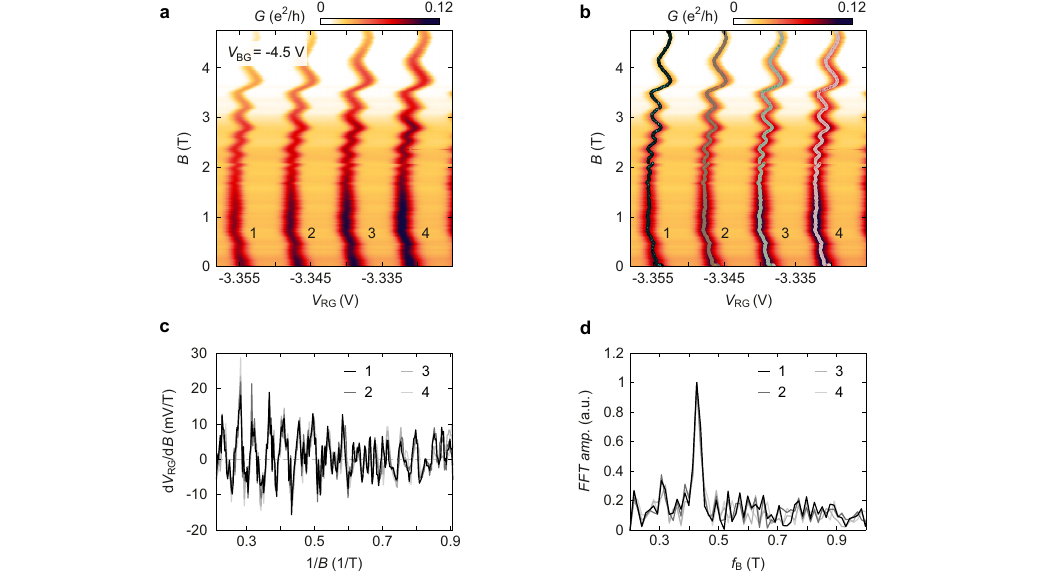}
\caption{\textbf{Proceeding of the frequency analysis}
(a) Tracked Coulomb resonances in the regime for $V_\mathrm{BG} = -4.5$~V. (b) Result of the peak finding algorithm. (c) Magnetic-field derivative of the peak positions $\mathrm{d}V_\mathrm{RG}/\mathrm{d}B$ as a function of inverse magnetic field $1/|B|$ for the individual traces. (d) Fast Fourier transform of the extracted oscillating signals shown in panel (c). The strong peak corresponds to the maximal frequency $F_B$.}\label{Sf_freq_analysis}
\end{figure}

We analyse the magnetic-field dependence of the tracked Coulomb resonances shown in Fig.~4 of the main manuscript and in \cref{Sf3} by first following the conductance maxima of the individual Coulomb resonances as a function of magnetic field via peak-finding using the Python package SciPy (\texttt{scipy.signal.find\_peaks}). 
We than fit a Lorentzian to the data points around the detected maxima to extract the peak position. 
In the next step, we calculate the numerical derivative along the magnetic field axis of the found peaks for each Coulomb resonance using NumPy's \texttt{numpy.gradient} method and plot them as a function of inverse magnetic field $1/B$.
After interpolation on an equally spaced $1/B$ axis we take the Fast Fourier Transform (FFT) using SciPy's \texttt{scipy.fft.fft} method.
We then take the average of the calculated gradients and the calculated FFT's and plot the data as a function of inverse magnetic field $1/B$ and the FFT frequency $f_B$, respectively.
In \cref{Sf_freq_analysis} we show this analysis exemplary for the regime recorded at a back gate voltage of $V_\mathrm{BG} = -4.5$~V (the final averaging of individual traces is not shown).

\clearpage
\section{Atomistic modeling of near-magic angle twisted bilayer graphene}

\paragraph{Atomic structure and lattice relaxation} To model the atomic and electronic structure of tBLG, we employ an atomistic modeling approach~\cite{trambly2010localization,Koshino2018Sep} and construct commensurate moiré unit cells with twist angles $\theta$ that are characterized by integer tuples ($n,m$)
\begin{equation}
\cos\theta=\frac{n^{2}+4 n m+m^{2}}{2\left(n^{2}+n m+m^{2}\right)}.
\end{equation}
For the simulations presented in the manuscript, we use $(n,m)=(26,27)$ corresponding to a twist angle of $\theta = 1.25^{\circ}$ which is closest to the experimentally extracted value of $\theta \approx 1.26^{\circ}$. 
The corresponding moiré unit cell contains $N=8428$ carbon atoms. 
The atomic positions are relaxed using classical force fields including a combination of AIREBO-Morse and Kolmogorov–Crespi potentials as used in Ref.~\cite{klebl2021importance}. \\

\paragraph{Electronic structure} The electronic structure of tBLG is captured within an atomistic tight-binding model that incorporates hopping processes between the carbon-$p_z$ orbitals~\cite{trambly2010localization, Koshino2018Sep} by distance-dependent Slater-Koster type hopping integrals $t(\bvec r)$
\begin{equation}
\hat H^{\text{TB}} = \sum_{\bvec{R} \bvec{R'}} \sum_{oo'} t(\bvec{R} + \bvec{r}_o - \bvec{R'} - \bvec{r}_{o'}) c_{\bvec{R}o}^{\dagger} c^{\phantom \dagger}_{\bvec{R'}o'}.
\label{eq:atomistic_tb_ham}
\end{equation}
The operator $c_{\bvec{R}o}^{(\dagger)}$ annihilates (creates) an electron in the $p_z$-orbital of the carbon atom at site $\bvec R + \bvec r_o$, where $\bvec R$ denotes a moir\'e lattice vector and $\bvec r_o$ denotes the positions of the carbon atom within the moir\'e unit cell. 
The Slater-Koster hopping terms are parameterized by
\begin{equation}
\begin{split}
t(\bvec d) &= t_{\parallel}(\bvec d) + t_{\bot}(\bvec d) \\
t_{\parallel}(\bvec d) &= -V_{pp \pi}^0\text{exp} \left(-\frac{d-a_{\text{cc}}}{\delta_0}\right) \left[1 - \left(\frac{d^z}{d} \right)^2 \right] \\
t_{\bot}(\bvec d) &= -V_{pp \sigma}^0\text{exp} \left(-\frac{d-d_0}{\delta_0}\right)\left[\frac{d^z}{d} \right]^2,
\end{split}
\label{eq:atomistic_sk}
\end{equation}
where $d = |\bvec d|$. 
The out-of-plane (in-plane) hopping amplitudes are chosen as $V^0_{pp \sigma} =  -0.48 \ \text{eV}$ ($V^0_{pp \pi} =  2.7 \ \text{eV}$), $d_0 = 1.362 \, a_0$ denotes the interlayer distance of graphite and $a_{cc} = a_0/\sqrt{3}$ is the nearest-neighbor distance between the carbon atoms in graphene ($a_0$ is the length of the graphene lattice vectors). 
The hopping integrals decay exponentially in real-space regularized by the factor $\delta_0 = 0.184 \, a_0$ and are thus truncated after 4th nearest-neighbor carbon atoms in numerical simulations. 
We further include the effect of external electrical fields by the operator
\begin{equation}
\hat H^{E} = \sum_{\bvec{R} o}  \varphi(\bvec R + \bvec r_o) c_{\bvec{R}o}^{\dagger} c^{\phantom \dagger}_{\bvec{R}o}\,,
\label{eq:atomistic_electrical}
\end{equation}
where $\varphi(\bvec R + \bvec r_o) = E [(\bvec R + \bvec r_o)\cdot \hat e_z]$ denotes the scalar potential associated to the external field. 
The displacement field extracted from experimental measurements, c.f. Section~\ref{sec:displacement_field_experimental}, already accounts for the encapsulating hBN and Al$_2$O$_3$ layers and we are therefore left with the internal screening induced by tBLG itself, which to first order is given by 
\begin{equation}
E = \frac{e D}{\varepsilon_0 \varepsilon_{\text{tBLG}}} + \mathcal{O}(\delta n)
\label{eq:internal_screening_tBLG}
\end{equation}
Estimates of the effective dielectric constant induced by screening from ordinary bilayer graphene range between $\varepsilon_{\text{BLG}} = 1 \dots 2$~\cite{icking2022transport,slizovskiy2021out}.
Due to the spatial inhomogeneity induced by the moir\'e pattern and the concomitant out-of-plane relaxation of the atomic structure that is particularly near the magic angle, this value may be further reduced compared to pristine bilayer graphene and we therefore assume $\varepsilon_{\text{tBLG}} = 1$, i.e. $E = D/\varepsilon_0$ in the following. 
However, we note that a more realistic model should additionally account for the occupation imbalance in the top (bottom) graphene layers $\delta n = n_T - n_B$ that intrinsically adds to the screening depending on the value of total density and displacement field and thus may render the relation $E(D)$ filling dependent~\cite{slizovskiy2021out}.\\

\paragraph{Valley quantum number in atomistic models}
Contrary to continuum model approaches~\cite{Bistritzer2011Jul,LopesdosSantos2007Dec, Shallcross2010Apr, SuarezMorell2010Sep}, the valley degree of freedom can not directly be accessed by the real-space tight-binding framework formulated above as it is related to an emergent aspect of the band structure in reciprocal space. 
However, the valley quantum number of the electronic states can be disentangled by using the Haldane-like valley operator proposed in Ref.~\cite{ramires2018electrically}
\begin{equation}
\hat V^{\nu} = \frac{i}{3 \sqrt{3}} \sum_{\llangle \bvec{R} \bvec{R'}} \sum_{oo' \rrangle} \eta^{\phantom z}_{\{\bvec R' o'  \bvec R o \}} \sigma^z_{\{\bvec R' o'  \bvec R o \}} c_{\bvec{R}'o'}^{\dagger} c^{\phantom \dagger}_{\bvec{R}o}\,,
\label{eq:atomistic_valley}
\end{equation}
where $\llangle \cdot \rrangle$ restricts the sum to the next-nearest (in-plane) neighbors on the honeycomb lattice, $\eta^{\phantom z}_{\{\bvec R' o'  \bvec R o \}} = \pm 1$ for clockwise and anti-clockwise hoppings and $\sigma^z_{\{\bvec R' o'  \bvec R o \}}$ is a Pauli matrix associated to the sublattice degrees of freedom of the original graphene sheets. 
The operator defined in Eq.~\eqref{eq:atomistic_valley} has expectation values $\pm 1$ in the valleys $K (K')$ and can thus be used to distinguish the valley-polarization of the electronic states obtained from the real-space tight-binding Hamiltonian $\hat H^{\text{TB}}$~\cite{ramires2018electrically}. \\

\paragraph{Hartree corrections}
Several studies~\cite{Guinea2018,goodwin2020hartree,Rademaker2019,Cea2019} have underlined the importance of long-ranged electron-electron interactions on the low-energy band structure of tBLG near the magic-angle. 
The impact of the Coulomb interaction can be captured to good extent within a self-consistent Hartree theory that leads to band renormalizations if the system is filled with electrons (holes). 
Ref.~\cite{goodwin2020hartree} demonstrated that for atomistic models of tBLG, the Hartree-induced band renormalizations can be described by a single-particle term of the form
\begin{equation}
\hat H^H = \sum_{\bvec{R} o}  V^H(\bvec R + \bvec r_o) c_{\bvec{R}o}^{\dagger} c^{\phantom \dagger}_{\bvec{R}o}\,.
\label{eq:atomistic_hartree}
\end{equation}
The Hartree potential $V^H$ is then to be solved self-consistently, however, the parametrization
\begin{equation}
V^H(\bvec r) = V_0 \nu \sum_{j} \cos(\bvec G_j \cdot \bvec r), 
\label{eq:atomistic_hartree_potential}
\end{equation}
was found to agree excellent for a wide range of twist angles and electronic fillings~\cite{Guinea2018,goodwin2020hartree,Rademaker2019,Cea2019}. 
Here, $\nu$ denotes the electronic filling $\nu = -4 \dots 4$ of the flat bands with respect to the charge neutrality point $(\nu=0)$ and $\bvec G_j$ are the three non-equivalent moir\'e reciprocal lattice vectors that differ by rotations around 120$^{\circ}$. 
The value of the Hartree potential $V_0$ was found to be $V_0 = 5$ meV for the unscreened Coulomb interaction, but will be reduced in the presence of external metallic gates or additional screening by the substrate~\cite{goodwin2020critical}.
\newpage

\section{Electronic structure under the influence of finite displacement fields}

\begin{figure}[!htbp]
\centering
\includegraphics[draft=false,keepaspectratio=true,clip,width=0.99\linewidth]{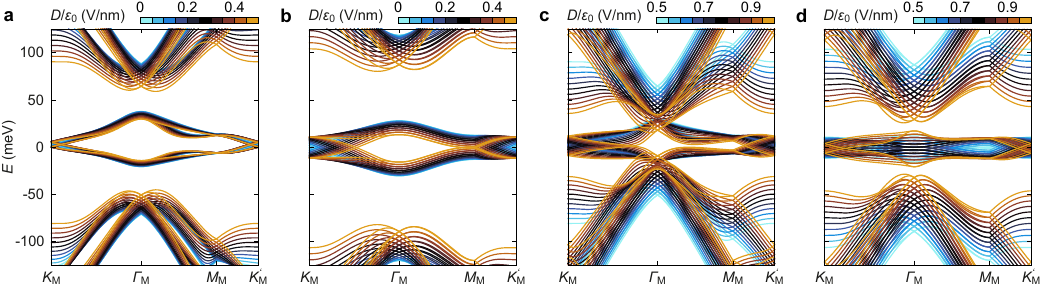}
\caption{\textbf{Influence of displacement fields on the band structure of tBLG.}
(a,c) Single-particle band structure of tBLG with a twist angle of $\theta = 1.25^\circ$ for different displacement fields $D/\varepsilon_0$ using the full tight-binding model as outlined in Eq.~\eqref{eq:atomistic_sk}.
(b,d) Single-particle band structure in the chirally symmetric (atomistic) model of tBLG for the same displacement fields shown in panels (a,c).
}\label{Sf_band_structure_D}
\end{figure}

\paragraph{Impact on flat electronic bands and single-particle gaps}
First, we study the band structure of tBLG under the influence of external displacement fields Eq.~\eqref{eq:atomistic_electrical} as shown in \cref{Sf_band_structure_D}a-d.
The presence of an interlayer bias lifts the valley degeneracy at the moir\'e Dirac points $K_M$ and $K_M'$ and causes an energetic splitting $\Delta_K$ between the energy doublets associated to the mini Dirac cones of each valley. 
Due to the presence of time-reversal symmetry, the band splitting induced by the displacement field is opposite in the two valley sectors. 
In particular, we find that the shape of the flat bands is merely altered for moderate displacement fields and the splitting at the $K_M$ and $K_M'$ points does not exceed $\Delta_K = 10$~meV as shown in \cref{Sf_band_structure_D}a.
Meanwhile the single-particle gap decreases continuously with increasing displacement field strength.
For large displacement fields $D>0.85$ V/nm, the band gap of tBLG closes and the flat bands hybridize with the remote valence and conduction bands as shown in \cref{Sf_band_structure_D}c. \\
\paragraph{Chirally symmetric (atomistic) model of tBLG}
Second, we suppress intra-sublattice interlayer tunneling such to effectively realize a chirally symmetric model of tBLG.  
In the atomistic tight-binding model Eq.~\eqref{eq:atomistic_tb_ham}, the chiral limit can be reached by explicitly labeling atoms in each graphene sheet according to the A (B) sublattice and suppressing interlayer tunneling between same sublattices.
In the process, the band gap of tBLG is enlarged to $\Delta_{\pm} \approx 100$ meV as demonstrated in \cref{Sf_band_structure_D}b. 
The enlarged band gap in the chiral limit was recently shown to explain optical excitations in near magic-angle devices~\cite{hesp2021observation} and could stem from the impact of the hBN substrate, electron-electron interactions or atomic relaxations favoring a charge redistribution to the AB (BA) regions within the moiré unit cell as a consequence of the strong electrical fields.
Opposite to the full tight-binding Hamiltonian, the shape of the flat bands undergoes major reconstructions as the band splitting at the moiré Dirac points becomes comparable to the flat band width.
Moreover, we find that for experimentally relevant displacement field strengths, the band gap never closes oppositely to the full tight-binding characterization shown in \cref{Sf_band_structure_D}a.
We note that the existence of a band gap in the areas underneath the finger gates (LG and RG) and split gates (SGs) is crucial to define the charge channel and to explain the pinch-off characteristics induced by the finger gates. 
Measuring the band gap in electrostatically defined tBLG in the presence of strong electrical fields could therefore validate theoretical models capturing the band structure of near magic-angle tBLG.\\

\paragraph{Extracting the quantum oscillation frequency from tight-binding calculations}

\begin{figure}[!htbp]
\centering
\includegraphics[draft=false,keepaspectratio=true,clip,width=0.9\linewidth]{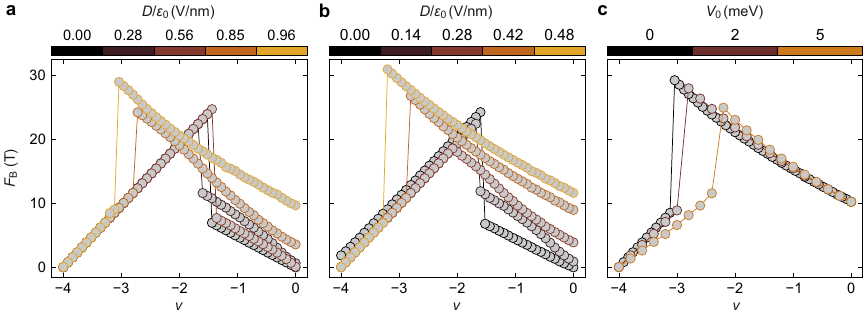}
\caption{\textbf{Theoretically predicted frequency tuning in 1.25$^{\circ}$ tBLG as function of the external displacement field $D$ and electronic filling $\nu = 0 \dots -4$ electrons per moir\'e unit cell.}
(a) Frequency tuning of the full tight-binding model as extracted from the size of the Fermi area $\mathcal{S}$ encircled by valley-exclusive orbits on the hole-doped side.
Without external displacement fields, the frequency $f_B$ tunes linearly near the CNP and the band edge, whereas the slope is inverted at the Lifshitz transition marking the crossover from electron-like orbits to hole-like orbits in the Fermi surface topology. 
Increasing the external displacement field $D$ shifts the Lifshitz transition to larger hole dopings.
At the same time, the quantum oscillation frequency takes a finite value at $\nu=0$ resulting from the charge carrier imbalance in the moir\'e Dirac cones as shown in \cref{Sf_Fermi_surface_TB}. 
(b) Same as in panel (a), but for the chirally symmetric tight-binding model of tBLG. 
While the qualitative features resemble the full tight-binding model, the required field strengths $D$ to (i) achieve a finite frequency at $\nu=0$ that is comparable with the experimental observations and (ii) displace the Lifshitz transition to large negative hole dopings are reduced by a factor of two.
(c) Influence of the the oscillations frequency in the presence of Hartree corrections induced by long-ranged Coulomb interactions for $D \approx 0.45$ V/nm in the chirally symmetric tight-binding model of tBLG.
}\label{Sf_frequency_electrical}
\end{figure}

\begin{figure}[!htbp]
\centering
\includegraphics[draft=false,keepaspectratio=true,clip,width=0.8\linewidth]{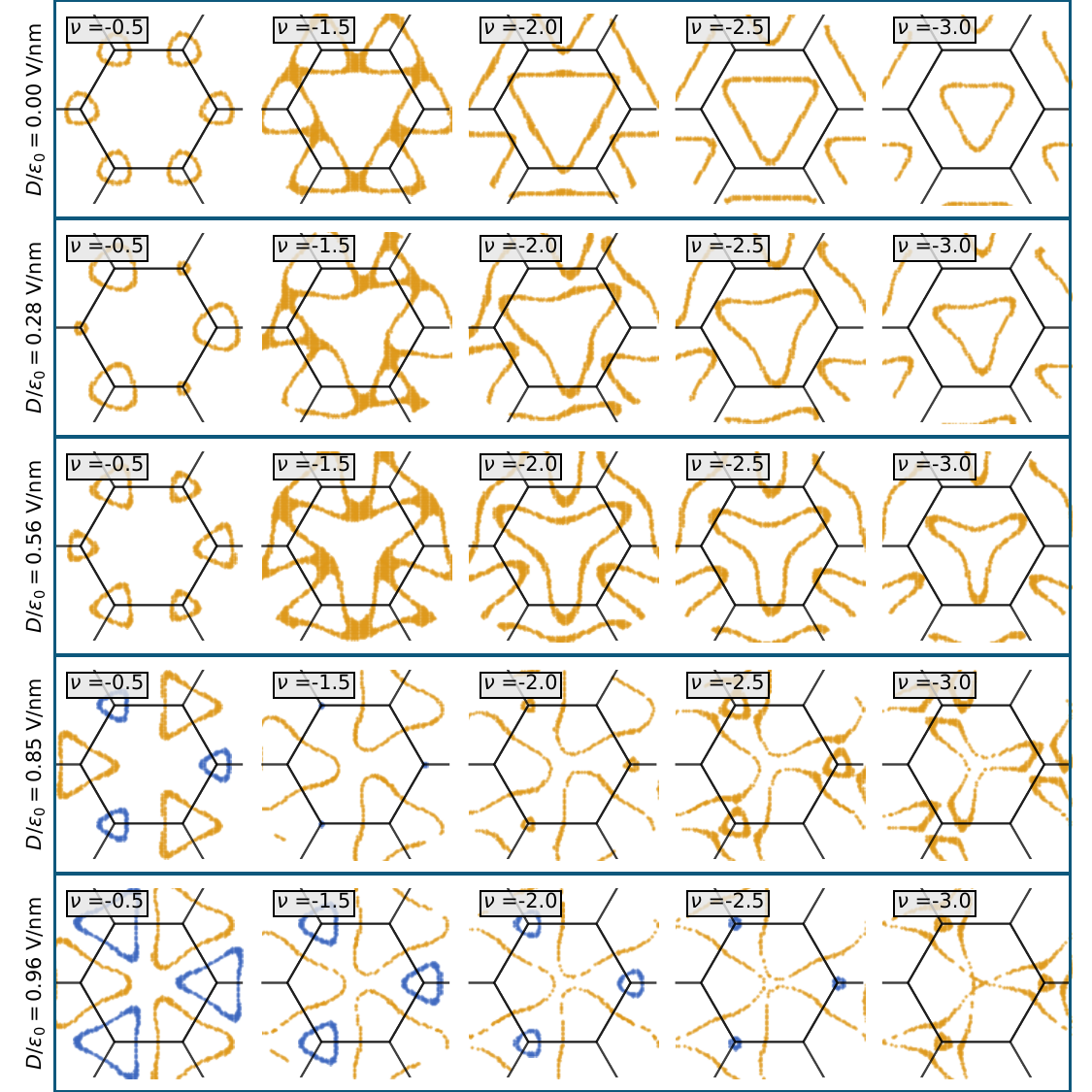}
\caption{\textbf{Fermi surface of 1.25$^{\circ}$ tBLG as function of the electronic filling $\nu$ and external displacement field $D$.}
At zero displacement field (upper row) the Fermi surface topology changes from valley-exclusive electron orbits centered around the moiré Dirac points $K_M^{(\prime)}$ to hole-like orbits centered around $\Gamma_M$. 
The Lifshitz transition occurs at fillings of $\nu = -1.5$.
Increasing the displacement field shifts the moir\'e Dirac cones causing a finite Fermi surface area at half-filling. 
As a consequence also the Lifshitz transition is shifted from $\nu = -1.5$ to larger hole dopings. 
At $D/\epsilon_0$, the Lifshitz transition is shifted down to $\nu = -3$.
The orange (blue) Fermi contours distinguish between the lower (upper) moiré flat band per valley contributing to the transport at the Fermi level.
}\label{Sf_Fermi_surface_TB}
\end{figure}

\begin{figure}[!htbp]
\centering
\includegraphics[draft=false,keepaspectratio=true,clip,width=0.8\linewidth]{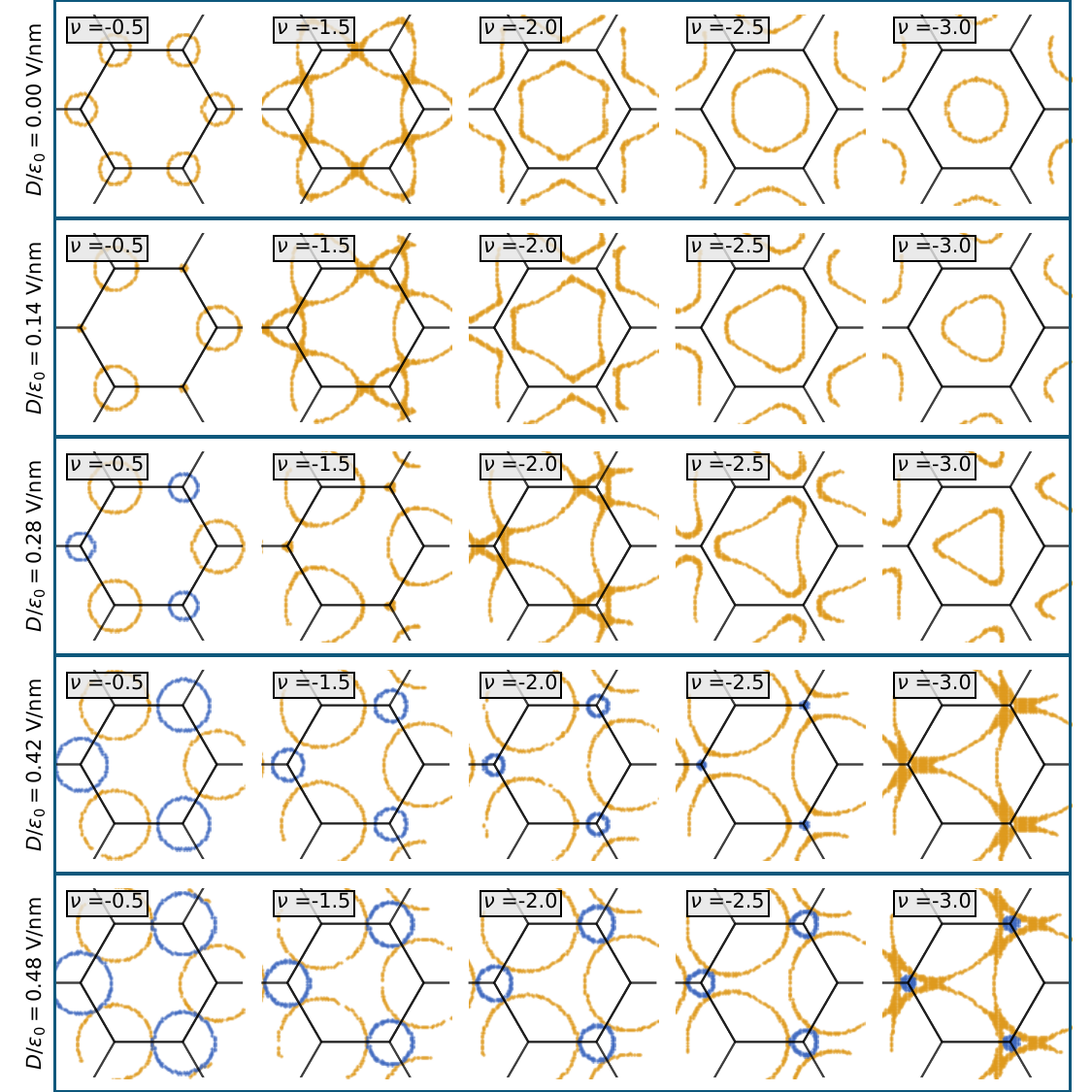}
\caption{\textbf{Fermi surface of 1.25$^{\circ}$ tBLG in the chirally symmetric TB model as function of the electronic filling $\nu$ and external displacement field $D$.}
At zero displacement field (upper row) the Fermi surface topology changes from valley-exclusive electron orbits centered around the moir\'e Dirac points $K_M^{(\prime)}$ to hole-like orbits centered around $\Gamma_M$. 
The Lifshitz transition occurs at fillings of $\nu = -1.5$ in analogue to the full TB model of tBLG shown in \cref{Sf_Fermi_surface_TB}.
Increasing the displacement field displaces the moir\'e Dirac cones in energy, which causes a finite Fermi surface area at half-filling. 
In addition, the potential difference between the layers shifts the Lifshitz transition to larger hole dopings. 
Contrary to the full TB model, the system retains a finite band gap separating the flat bands and the remote valence (conduction) bands, c.f. \cref{Sf_band_structure_D}.
The orange (blue) Fermi contours distinguish between the lower (upper) moir\'e flat band per valley contributing to the transport at the Fermi level.
}\label{Sf_Fermi_surface_chiral}
\end{figure}

To connect to the experimentally extracted frequency $F_B$ of the $1/|B|$ oscillations under an external magnetic field, we exploit the Onsager relation
$\mathcal{S} = 2 \pi e F_B/ \hbar$. 
The Fermi surface area $\mathcal{S}$ is extracted by sampling the Brillouin zone with an equidistant mesh of $N_{\bvec k} = 60 \times 60$ points and measuring the (finite) connected area spanned by electron-like (hole-like) states. 
Close to the magic-angle, intervalley scattering is suppressed~\cite{Cao2016Sep} and we therefore only consider the area encircled by valley-exclusive orbits, which is achieved by labeling electronic states in the flat band manifold according to their valley quantum number by virtue of Eq.~\eqref{eq:atomistic_valley}. 
The Fermi surfaces for different electronic fillings and displacement field strengths for the full (chirally symmetric) tight-binding model of tBLG are shown in \cref{Sf_Fermi_surface_TB} (\cref{Sf_Fermi_surface_chiral}).
At zero electrical field, the Fermi surface changes from valley-exclusive, hole-like orbits that encircle the moir\'e Dirac points $K_\mathrm{M}^{(\prime)}$ to electron-like orbits centered around $\Gamma_\mathrm{M}$ as shown in \cref{Sf_Fermi_surface_TB} (upper row).
Increasing the electrical field ($D > 0$~V/nm) leads to an energetic splitting of the moir\'e Dirac points $K_\mathrm{M}^{(\prime)}$ such that the upper and lower moir\'e flat band contribute to the Fermi surface as indicated by the blue and orange contours respectively.
The induced occupation imbalance causes a finite Fermi surface area $\mathcal{S}$ and a finite quantum oscillation frequency at $\nu = 0$ and obviates the Lifshitz transition to larger hole dopings $\nu < -3$ in agreement with the experimental measurements.
In the chirally symmetric tight-binding model \cref{Sf_Fermi_surface_chiral}, the phenomenology is similar, however, the required electrical field strengths needed to cause a finite Fermi surface area at the CNP roughly differ by a factor of two. \\
Finally, we extract the frequency $F_B$ of the quantum oscillations as function of the electronic filling  for the full (chirally symmetric) TB model as shown in \cref{Sf_frequency_electrical}. 
In agreement with the discussion of the Fermi surface of tBLG at different electrical field strengths, there are two major insights that align with the experimental observations: (i) the Lifshitz transition is shifted to larger hole dopings for strong electrical fields and (ii) the presence of external fields causes a finite oscillations frequency already at $\nu=0$.
We note that the offset of the quantum oscillation frequency at the CNP can only be explained in chirally symmetric model of tBLG for realistic displacement field strengths while retaining the single-particle band gap. \\

\paragraph{Influence of Hartree corrections}

As the experimentally determined twist angle of the sample is close to the magic angle of tBLG, we further include the effect of long-ranged Coulomb interactions by accounting for the Hartree potential $V^H$ derived in Eq.~\eqref{eq:atomistic_hartree_potential}.
The strength of the Hartree potential will in general depend on the dielectric environment and particularly the geometry of the gate architecture~\cite{Goodwin2020screening} such that we treat the magnitude of the Hartree potential $V_0$ as free parameter as shown in \cref{Sf_frequency_electrical}~(c).
Our analysis for the chirally symmetric tight-binding model of tBLG suggests that the frequency tuning observed in the tBLG charge island does not change qualitatively for small values of the Hartree potential $V_0 \leq 2$ meV as expected for moderate to strong screening of the long-ranged Coulomb interaction. 
However, using the unscreened value $V_0 = 5$ meV~\cite{goodwin2020hartree}, shifts the Lifshitz transition back to smaller hole dopings $\nu = -2$ in contrast to the experimental findings. 
These observations suggests that the gating architecture and the proximity of the tBLG sample to the finger and split gates causes significant screening of the long-ranged Coulomb interactions such to suppress the relevance of Hartree corrections to the electronic structure. 
This would be in line with the absence of the correlated insulating state at $\nu=-2$ in the constriction and could point towards the importance of long-ranged Coulomb interactions to the formation of the correlated insulating state observable in the bulk transport.

\newpage

\clearpage

\section{Displacement field model for the gating architecture}\label{sec:displacement_field_experimental}
Our gating architecture yields to a complex situation of different displacement fields present in the individual confinement regimes. 
These displacement fields are defined by the two split gates SG1 and SG2 and the two finger gates LG and RG together with the graphite BG in each case. 
The displacement field between the back gate and the split gates is given by
\begin{equation}
    D_\mathrm{SG1/SG2} = \frac{e}{2}\left[\alpha_\mathrm{BG} (V_\mathrm{BG} - V_\mathrm{BG, off})  - \alpha_\mathrm{SG}(V_\mathrm{SG} -V_\mathrm{SG,off}) \right].
\label{eq:displacement_bg}
\end{equation}
Analogously, the displacement fields between the finger gates and the back gate reads
\begin{equation}
    D_\mathrm{LG/RG} = \frac{e}{2}\left[\alpha_\mathrm{BG} (V_\mathrm{BG} - V_\mathrm{BG, off})  - \alpha_\mathrm{FG}(V_\mathrm{FG} -V_\mathrm{FG,off}) \right],
\label{eq:displacement_fg}
\end{equation}
with
\begin{align}
    \alpha_\mathrm{FG} = \frac{\varepsilon_0}{2e} \left[\frac{\varepsilon_\mathrm{Al_2O_3}}{d_\mathrm{Al_2O_3}} + \frac{\varepsilon_\mathrm{hBN}}{d_\mathrm{t}} \right].
\end{align}
We note, that with this definition of the displacement field  \cite{icking2022transport} the quantity $D/\varepsilon_0$ is measured in units of $\mathrm{V/nm}$.
For the displacement fields between the split gates and the back gate, we reach values ranging from $\approx -0.23 \, \mathrm{V/nm}$ and $\approx 0.15 \, \mathrm{V/nm}$ while we reach values for the displacement fields between the finger gates and the back gate ranging from  $\approx -0.48 \, \mathrm{V/nm}$ and $\approx 0.80 \, \mathrm{V/nm}$.
\clearpage